\documentclass[12pt,a4paper]{article}
\usepackage{amssymb}
\def\pa{\partial}

\def\dfrac #1#2{\displaystyle{\frac{#1}{#2}}}

\newcommand{\initiate}{\setcounter{equation}{0}}

\newcommand{\beq}{\begin{equation}}
\newcommand{\eeq}{\end{equation}}
\newcommand\be{\begin{equation} }
\newcommand\bea{\begin{eqnarray}}
\newcommand\ee{\end{equation}}
\newcommand\eea{\end{eqnarray}}

\def\Tr{{\rm Tr}\,}

\def\endtitle{\par\end{quotation}\vskip3.5in minus2.3in\newpage}

\def\a{\alpha}       \def\b{\beta}
         \def\d{\delta}
\def\e{\epsilon}     
\def\g{\gamma}       
        
       \def\l{\lambda}
\def\m{\mu}          \def\n{\nu}
       
       \def\r{\rho}
\def\s{\sigma}

         \def\G{\Gamma}

\def\ca{{\cal A}}

\def\ci{{\cal I}}      
      \def\cl{{\cal L}}

\begin{document}

\title{Renormalizability of noncommutative $SU(N)$ gauge theory}

\author{Maja Buri\' c\footnote{E-mail: majab@phy.bg.ac.yu}, Du\v sko Latas\footnote{E-mail: latas@phy.bg.ac.yu}
 and Voja Radovanovi\' c\footnote{E-mail: rvoja@phy.bg.ac.yu} \\
Faculty of Physics, University of Belgrade\\
P.O.Box 368, 11001 Belgrade, Serbia and Montenegro}

\date{}

\maketitle

\abstract{We analyze the renormalizability properties of pure
gauge noncommutative $SU(N)$ theory in the $\theta$-expanded approach. We find
that the theory is one-loop renormalizable to first order in
$\theta$.}

\vfill \noindent  \eject
\parskip 4pt plus2pt minus2pt

\initiate \section{Introduction}

The early motivation to introduce noncommutativity of coordinates
was the search for invariant regulator in field theory. Many other
reasons for studying noncommutative manifolds have appeared since;
still, the renormalizability of field theories, on
noncommutative Minkowski  space  for example, has not been established. The aim
of this paper is to add further results to this discussion.

To  introduce the problem and the notation, we give a couple of
basic definitions. Noncommutative Minkowski space is the algebra
generated by  coordinates $\hat x^\mu$, $\mu = 0,1,2,3$, which
obey the commutation relations of canonical type:
\begin{equation}
[\hat x^\mu,\hat x^\nu ] =i\theta^{\mu\nu}={\rm const}.                 \label{Mink}
\end{equation}
The algebra (\ref{Mink}) can be represented by the algebra of functions on commutative four-dimensional
 manifold with the Moyal-Weyl
product as multiplication. The latter  is defined as
\begin{equation}
\label{moyal} \phi (x)\star \chi (x) =
      e^{\frac{i}{2}\,\theta^{\m\n}\frac{\pa}{\pa x^\m}\frac{\pa}{ \pa
      y^\n}}\phi (x)\chi (y)|_{y\to x}\ .
\end{equation}
The fields  are functions of coordinates; for example, the vector
potential  $\hat A_\mu(x)$ of the gauge group $U(N)$ is, in this
context,  defined by
\begin{equation}
\hat A_\mu ( x) = \hat A_\mu^A(x){\bf t}^A, \label{A}
\end{equation}
where   ${\bf t}^A$ are ordinary $N\times N$ matrices, generators
of  $U(N)$. We denote the generators of $SU(N)$ by $t^a$:  the
capital letters denote the $U(N)$ indices, while the small letters denote
 the $SU(N)$ indices,  ${\bf t}^A\in \{ I, t^a\} $.  The field strength transforms in the adjoint representation; it is
given by
\begin{equation}
\hat F_{\mu\nu} = \pa _\mu\hat A_\nu - \pa _\nu\hat A_\mu -i(\hat
A_\mu\star\hat A_\nu -\hat A_\nu\star\hat A_\mu ). \label{F}
\end{equation}
The generators of $SU(N)$ satisfy \beq [t^a,t^b]=if^{abc}t^c, \qquad
\{t^a,t^b\}=d^{abc}t^c ,\eeq
 $f^{abc}$ are the structure constants of $SU(N)$, $d^{abc}$ are
the symmetric symbols, $d^{abc}=\Tr \{t^a,t^b\}t^c$: we use the
normalization $\Tr(t^at^b)=\d^{ab}$. For  gauge groups different
from $U(N)$ the commutator term in (\ref{F}) does not take values
in the Lie algebra of the group, as the $\star$-product is not
commutative. It is  clear therefore that not all gauge groups can
be realized on the  space (\ref{Mink}) in the described manner.

As the gauge fields are represented by functions on ${\bf R}^4$,
the integration and  the action can be defined
straightforwardly. The action for the pure gauge theory reads
\begin{equation}
S=-\frac{1}{4}\,\Tr \int d^4x\hat F_{\m\n}\star\hat F^{\m\n} =-\frac{1}{4}\,
\Tr\int d^4x\hat F_{\m\n}\hat F^{\m\n} ,
\label{action}
\end{equation}
the last equality is valid due to  properties of the Moyal-Weyl
multiplication. One can easily include matter fields. This
basically concludes the definition of the classical  theory.
Compared to the commutative gauge theory, it has  new, specific
features. For quantization, one usually takes $\theta^{0i}=0$ as
in that case  there are no temporal derivatives of higher order in
the lagrangian.  This means that the generalized momenta are the
same as in the classical theory and  the momentum dependence in
the path integral is Gaussian. Stated differently, $\theta^{0i}
=0$ provides the unitarity. In perturbation expansion  Feynman
rules get modified in accordance with the definition of
$\star$-multiplication.  The most important result concerning
renormalization is the mixing of ultraviolet and infrared
divergencies: in higher-order diagrams they are entangled in such
a way that no efficient renormalization procedure can be defined.
This phenomenon is typical for noncommutative theories and has
been thoroughly discussed for $\phi^4$ theory, \cite{phi4}. See
also the alternative heat-kernel derivation in \cite{Gay}. The
existence of gauge symmetry does not remove UV/IR mixing: for more
detailed analysis for $U(1)$ we refer to \cite{U1}, for nonabelian
theories to  \cite{nonab}. Consequently, the status of
renormalizability of noncommutative gauge theories is the same as
for noncommutative scalar field theory: in general,
 the theories are not renormalizable.

A different representation of noncommutative  gauge theories on
commutative ${\bf R^4}$ was developed in \cite{thetaexp}. Its
main idea is to enlarge  the basis of the algebra to the
enveloping algebra of the group. Although only the infinitesimal  gauge
transformations can be defined in this way, the construction
goes for arbitrary gauge groups and their tensor products.  As it
was shown in \cite{thetaexp}, every physical field
can written as a formal expansion  in noncommutativity parameter
$\theta^{\m\n}$; the leading term in the expansion of noncommutative field is
its commutative counterpart. Therefore, for $\theta^{\m\n}= 0$
noncommutative theory reduces to the usual, commutative one: the
representation is a deformation of the commutative theory. This
fact can also be related to the Seiberg-Witten result \cite{SW}
that classically noncommutative and commutative gauge theories are
equivalent. The  expansions of the vector potential and the field
strength to linear order in $\theta $ read
\bea
&&\hat A_\r(x) =A_\r(x) -\frac 14 \,\theta ^{\m\n}\left\{ A_\m(x),
\pa _\n A_\r(x) +F_{\n \r}(x)\right\}
+\dots\label{expansion}\\
&&\hat F_{\r\s}(x) =   F_{\r\s}(x) +\frac{1}2{}\theta^{\m\n} \{
F_{\m\r}(x),F_{\n\s}(x)\} -\frac{1}{4}\theta^{\m\n}\{ A _\m (x),
(\pa_\n +D_\n )F_{\r\s} (x)\}+\dots , \nonumber
\eea
and, as it was noted in \cite{a}, are not unique. The gauge field
action (\ref{action}) can be expanded, too. The parameter
$\theta$ is  treated as a coupling constant and thus
propagators of all fields are defined as in the commutative theory. It can be
shown that each of the interaction terms  is invariant to
commutative gauge transformations; noncommutative gauge symmetry
is recorded only in the complete sum (\ref{expansion}).

Obviously, a drawback of the `$\theta$-expanded' approach is that
the results are necessarily expressed in powers of $\theta$ and
calculated to a certain order: in practice, at most to second
order. Thus one cannot obtain results which are nonanalytic in
$\theta$: and UV/IR mixing is an effect proportional to
$\theta^{-1}$.  On the other hand, it does make very good sense to
expand in $\theta$ if one wants to compare with the experiment: if
exists, noncommutativity  is very small.  The question of
renormalizability   can be approached as well; in particular,
negative results which one might obtain can be regarded
conclusive. Let us stress that, in principle, there is no reason
to expect to get the same results as for the nonexpanded theory,
as the two representations of the gauge symmetry are different; in
fact, $SU(N)$ cannot even be defined in the nonexpanded
representation. It was shown in \cite{w,mv,mv1} that $U(1)$ and
$SU(2)$ gauge theories with fermions are not renormalizable: the
divergencies which cannot be removed exist both at $\theta$-linear
and $\theta$-quadratic level.  The results concerning  pure gauge
theories on the other hand, are somewhat partial: for example, the
photon propagator in $U(1)$ theory is renormalizable in a
generalized sense, to linear and possibly to all orders,
\cite{u1}. Similar statement holds for the gluon propagator in
$SU(2)$ to linear order, \cite{mv1}. In this paper we investigate
renormalizability of the first order-corrected $SU(N)$  theory: as
we shall see from the form of divergencies, theory is in this
order  renormalizable. This result, we think, keeps the discussion
on the renormalizability of field theories on noncommutative
Minkowski space open.

The plan of the paper is the following. In the next section we
expand the classical action to second order in quantum fields and
do the path integral quantization. In the third section
we present the result of the calculation of divergencies in
the one-loop effective action and we show that the action can be
renormalized in a very simple way. A summary of the results is given in the
concluding section.

\initiate
\section{Expansion in background fields}

The classical action for the  gauge field expanded to first order
in noncommutativity  is \cite{thetaexp}:
{\setlength\arraycolsep{2pt}
\begin{eqnarray}
S_{cl} &=&-\frac{1}{4}\Tr\int d^4x\hat F_{\m\n}\star\hat F^{\m\n} \nonumber\\
&=&\Tr\int d^4x\left( -\frac{1}{4}F_{\m\n}F^{\m\n}+\frac{1}{8}
\theta^{\m\n}F_{\m\n}F_{\r\s}F^{\r\s}-\frac{1}{2}\theta^{\m\n}
F_{\m\r}F_{\n\s}F^{\r\s} \right)\label{actionexp} \\
&=&\int d^4x\left( -\frac{1}{4}F_{\m\n }^aF^{\m\n a}+
\frac{1}{16}\theta^{\m\n}d^{abc}(F_{\m\n}^aF_{\r\s}^bF^{\r\s c}
-4F_{\m\r}^aF_{\n\s}^bF^{\r\s c}) \right).\nonumber
\end{eqnarray}}
We are dealing with the gauge group $SU(N)$: $a,b,c=1,\dots
,N^2-1$ are the group indices. The group metric is Euclidean so
there is no need to distinguish between upper and lower indices.
The gauge fields are assumed to be in the adjoint representation;
we will comment   further on this restriction. We quantize the
theory by functional integration of the  vector potential. The
integration is done around the classical configuration, i.e., we
use the background field method to find the effective action. The
main contribution to the integral is given by the Gaussian
integral: to find it, technically, we  need to write  the vector
potential  in the action  as a sum of the background  part
$A_\m^a$ and the quantum fluctuation $\ca _\m^a$, and  find the
part  quadratic in $\ca ^a_\mu$. By this we determine the second
functional derivative of the classical action  with respect to the
vector potential. In the saddle-point approximation, the result of
the functional integration is
\begin{equation}
\Gamma [A_\mu^a] = S[A_\m ^a]+\frac{i}{2}\log \det
S^{(2)}[A_\m^a].
\end{equation}
We are dealing with the gauge symmetry and therefore the gauge fixing term
has to be included  in the action as well:
\begin{equation}
S = S_{cl} +S_{gf}, \qquad S_{gf} =
-\frac{1}{2} \int d^4x(D_\m \ca^{\m a})^2,
\end{equation}
with $D_\m\ca _\n ^a = \pa _\m\ca _\n ^a +f^{abc}A_\m ^b\ca _\n ^c
$. The one-loop effective action,
\begin{equation}
\Gamma ^{1}[A_\m^a] =\frac{i}{2}\log\det
S^{(2)}[A_\m^a]=\frac{i}{2}\Tr\log S^{(2)}[A_\m^a],                  \label{gama1}
\end{equation}
can be obtained in the usual way, by perturbative expansion of the
logarithm. As we  already explained the method in details in
\cite{mv},  we discuss here only the points specific to the present
calculation.

In the case we are studying, the quadratic part of the action is
very complicated. Very involved too are the operator traces
whose divergent parts we calculate by dimensional regularization.
It is necessary therefore to develop a strategy already at this
stage.  We will calculate the  divergencies at the special,
constant  value of the classical vector potential,
\begin{equation}
A_\m ^a ={\rm const}.               \label{Aconst}
\end{equation}
At the end of the calculation the full expressions will be
restored from covariance, replacing $f^{abc}A_\m ^bA_\n ^c$ by $
F_{\m\n}^a $. The idea is not new; for a similar derivation see
\cite{weinberg}. The use of background field method guarantees the
covariance \cite{ps}, as doing the path integral, we fix the local
symmetry of the quantum field $\ca_\m^a$ while the
 gauge symmetry of the background field $A_\m^a$ is manifestly preserved.
  Having in mind the dimension-regularization formulae, we
see that in fact it is an advantage to deal with nonabelian
theory, as the terms without derivatives  are by far the simplest.
At the moment, the assumption (\ref{Aconst}) means just that
derivatives commute with $A_\m^a$ in $S^{(2)}[A_\mu ^a]$.

Extracting the quadratic part of the action is a straightforward
but tedious  calculation. The result has the form
\beq
S^{(2)}=\frac12\ca^a_\a\left[g^{\a\b}\d_{ab}\Box+(N_1+N_2+T_2+T_3+T_4)^{\a
a ,\b b}\right]\ca^b_\b .                              \label{S2}
\eeq
The operators $N_1$ and $N_2$ originate from the commutative
action, while $T_2$, $T_3$ and $T_4$ denote  interaction terms
linear in $\theta$. The index of the operator  indicates
the number of background fields $A_\mu^a$ which it contains. The operators are given by
 \beq (N_1)^{\a a ,\b b}=-2f^{abc}A_\m^cg^{\a\b} \pa^\m= -2i
(A_\m )_{ab}g^{\a\b}\pa^\m,
 \eeq
where we introduced the matrix notation
\beq
(A_\m)_{ab} = -if^{abc}
A_\m^c = A_\m^c(T^c)_{ab},
 \eeq
because  the structure constants are the matrix elements of
generators in the adjoint representation. The same notation will
be used for the field strengths as it is very useful. For example,
 \beq
(A_1\dots A_{2n})_{ab} = (A_{2n}\dots A_1)_{ba},
\quad (A_1\dots A_{2n+1})_{ab} = -(A_{2n+1}\dots A_1)_{ba}.
\eeq
The rest of the vertices in (\ref{S2}) read
\beq
(N_2)^{ab}_{\a\b}=-2f^{cab}F_{\a\b}^{c}-(A_\m A^\m)^{ab}g_{\a\b},
\eeq {\setlength\arraycolsep{2pt}\bea (T_2)^{ab}_{\a\b}&=& \frac14
d^{abc} \left[-2\theta_{\r\a}F^c_{\m\b}\pa^\m\pa^\r- 2
\theta_{\r\b} F^c_{\m\a} \pa^\m \pa^\r - \theta^{\r\s} F^c_{\r\s}
g_{\a\b} \Box+ \theta^{\r\s} F^c_{\r\s} \pa_\a \pa_\b
\right.\nonumber\\
&+& 2{\theta^\r}_{\a}F^c_{\m\r }\pa^\m\pa_\b +2{\theta^\r}_{\b}
F^c_{\m\r} \pa^\m \pa_\a -2 {\theta^\r}_{\b} F^c_{\a\r} \Box-2
{\theta^\r}_{\a} F^c_{\b\r} \Box
-4\theta^{\r\s} F^c_{\m\r} g_{\a\b} \pa^\m \pa_\s\nonumber\\
&+& \left. 2\theta^{\r\s} F^c_{\b\r} \pa_\a \pa_\s +
2\theta^{\r\s} F^c_{\a\r} \pa_\b \pa_\s - 2 \theta_{\a\s}
F^c_{\m\b} \pa^\m \pa^\s - 2 {\theta^\r}_{\b}
F^c_{\a\n}\pa^\n\pa_\r \right] , \nonumber \eea}
{\setlength\arraycolsep{2pt}\bea (T_3)^{ab}_{\a\b}&=
&-\frac{i}{4}d^{acd}\left[2\theta_{\r\a}F_{\m\b}^d(A^\m)_{bc}\pa^\r
+ 2\theta_{\r\b}F_{\m\a}^d(A^\r)_{bc}\pa^\m\right. \nonumber\\
&+& \theta^{\r\s}g_{\a\b}F_{\r\s}^d(A^\m)_{bc}\pa_\m
-\theta^{\r\s}F_{\r\s}^d(A_\a)_{bc}\pa_\b \nonumber\\
&-& 2 {\theta^{\r}}_{\a} F_{\m\r}^d (A^\m)_{bc}\pa_\b + 2
{\theta^{\r}}_{\a} F_{\b\r}^d (A^\n)_{bc}\pa_\n\nonumber\\
&+&2 \theta^{\r\s} g_{\a\b} F_{\m\r}^d (A^\m)_{bc} \pa_\s - 2
\theta^{\r\s} F_{\b\r}^d (A_\a)_{bc} \pa_\s -2
{\theta^{\r}}_{\b} F_{\m\r}^d (A^\a)_{bc} \pa^\m \nonumber\\
&+& 2 {\theta^{\r}}_{\b} F_{\a\r}^d (A^\n)_{bc} \pa_\n + 2
\theta^{\r\s} g_{\a\b} F_{\m\r}^d (A^\s)_{bc} \pa^\m
- 2 \theta^{\r\s} F_{\a\r}^d (A_\s)_{bc} \pa^\b \nonumber\\
&-& 2 \theta_{\a\b} F_{\m\n}^d (A^\n)_{bc} \pa^\m +2 \theta_{\a\s}
F_{\m\b}^d (A^\s)_{bc} \pa^\m
+2 \theta_{\r\b} F_{\a\n}^d (A^\n)_{bc} \pa^\r \nonumber\\
&-&\left. 2 \theta^{\r\s} F_{\a\b}^d
(A_\s)_{bc}\pa_\r-(a\leftrightarrow b,\a
\leftrightarrow\b)\right], \nonumber \eea} and
{\setlength\arraycolsep{2pt}\bea (T_4)^{ab}_{\a\b}&=& \frac18
d^{ced} \left[-4\theta_{\r\a}F_{\m\b}^e(A^\r)_{ad}(A^\m)_{bc}
-\theta^{\r\s}g_{\a\b}F_{\r\s}^e(A^\m)_{ad}(A_\m)_{bc}\right. \nonumber\\
&+&\theta^{\r\s}F_{\r\s}^e(A_\b)_{da}(A_\a)_{cb} +4{\theta^{\r}}_{\a}
F_{\m\r}^e(A_\b)_{ad}(A^\m)_{bc} \nonumber\\
&-&4 {\theta^{\r}}_{\a} F_{\b\r}^e (A^\n)_{ad} (A_\n)_{bc} -
4\theta^{\r\s}g_{\a\b}F_{\m\r}^e(A_\s)_{ad}(A^\m)_{bc}\nonumber\\
&+& 4 \theta^{\r\s}F_{\b\r}^e(A^\s)_{ad}(A_\a)_{bc} + 2
\theta_{\a\b} F_{\m\n}^e (A^\m)_{ad}(A^\n)_{bc}\nonumber\\
&-& 2 \theta_{\a\r} F_{\m\b}^e (A^\m)_{ad}(A^\r)_{bc}
- 2 \theta_{\r\b}F_{\a\n}^e(A^\r)_{ad}(A^\n)_{bc}\nonumber\\
&+& \left.2\theta_{\r\s} F_{\a\b}^e (A^\r)_{ad}(A^\s)_{bc} - 4
{\theta^{\r}}_{\b} f^{cab}F_{\m\r}^d {F^{\m}_{\ \a}}^e +
(a\leftrightarrow b,\a \leftrightarrow\b) \right]\nonumber \\
&+&\frac18 d^{ced} f^{cab} \left[2\theta^{\r\s}F_{\r\s}^d
F_{\a\b}^e +
\theta_{\a\b} F_{\m\n}^d F^{\m\n e} \right]\nonumber\\
&-& \frac14\theta^{\r\s}f^{cab}d^{cde}(F_{\a\r}^dF_{\b\s}^e
-F_{\b\r}^dF_{\a\s}^e). \nonumber \eea}
As already noted, they
are written under the assumption (\ref{Aconst}).

\initiate
\section{Divergences}

From (\ref{S2}) we read off  the second functional derivative of
the action:
\beq (S^{(2)}[A_\mu^a])^{\a a,\b
b}=g^{\a\b}\d_{ab}\Box+(N_1+N_2+T_2+T_3+T_4)^{\a a ,\b b} .
\label{S2mat} \eeq
In order to calculate the effective action
(\ref{gama1}), we expand the logarithm. Denoting $ \ci ^{\a a,\b
b} = g^{\a\b}\d_{ab} $, we write
{\setlength\arraycolsep{2pt}
\begin{eqnarray}\label{sum} \Tr\log S^{(2)}[A_\mu^a]& =&\Tr\log \ci\Box
+\Tr\log\left(\ci +\Box^{-1} (N_1+N_2+T_2+T_3+T_4)\right)
\nonumber \\ &=& \sum_{n=1}^\infty\frac{(-1)^{n+1}}{n}\ \Tr
\left(\Box^{-1}(N_1+N_2+T_2+T_3+T_4)\right)^n,
\end{eqnarray}} \noindent
where in the last equality  the
infinite normalization constant is neglected.
The terms which contribute to the divergent one-loop effective
action in  zero-th order are
\begin{equation}
\G^{0}=\frac{i}{2}\left(-\frac12\Tr(\Box^{-1}N_2)^2
+\Tr((\Box^{-1}N_1)^2\Box^{-1}N_2)
-\frac14\Tr(\Box^{-1}N_1)^4\right) ;
\end{equation}
so the divergent part is
\begin{equation}
\G ^0|_{\rm div} = \frac{5N}{3(4\pi )^2\epsilon}\int d^4x
F_{\m\n}^aF^{\m\n a}.            \label{g0}
\end{equation}
However, the ghost contribution  should also be taken into
account. It is derived from the ghost action introduced as in
\cite{mv1,thoft,ps}
\beq
 \G_{gh}|_{\rm div}=
\frac{N}{6(4\pi )^2\epsilon}\int d^4x F_{\m\n}^aF^{\m\n a} .
\label{ggh} \eeq Thus the sum of (\ref{g0}) and (\ref{ggh})
constitutes the standard result of the commutative theory. Our
goal is to calculate the $\theta$-linear divergencies. As
$\theta^{\m\n} $ has the length dimension $2$ and $F^{\m\n }$ is
of dimension $-2$, the terms of the following types are possible:
$\theta F^3, $ $\theta (DF)^2,$ $ \theta \epsilon F^3,$ $\theta
\epsilon (DF)^2$. All of them have one $\theta$ and six $A$'s:
thus from  (\ref{sum}) we need to extract the terms which contain
one of the vertices $T_i$ and have the sum of indices equal to 6.
The trace to be calculated is \bea \Gamma ^1|_{\rm
div}&=&\frac{i}{2}\left( -\Tr[\Box^{-1}N_2 \Box^{-1}T_4]
+\Tr[(\Box^{-1}N_1)^2 \Box^{-1} T_4] + \Tr[(\Box^{-1} N_2)^2
\Box^{-1}T_2] \right. \nonumber \\
&+& \Tr[\Box^{-1} N_1 \Box^{-1} N_2 \Box^{-1} T_3] + \Tr[\Box^{-1}
N_2
\Box^{-1} N_1 \Box^{-1} T_3] - \Tr[(\Box^{-1} N_1)^3 T_3] \nonumber\\
&-& \Tr[(\Box^{-1} N_1)^2 \Box^{-1}N_2 \Box^{-1}T_2] -
\Tr[\Box^{-1} N_2
(\Box^{-1} N_1)^2 \Box^{-1}T_2] \nonumber \\
&-& \left.\Tr[\Box^{-1} N_1 \Box^{-1}N_2 \Box^{-1}N_1
\Box^{-1}T_2] + \Tr[(\Box^{-1} N_1)^4 \Box^{-1} T_2] \right) .
 \eea
In fact, using Bianchi identities it can easily be seen that all
possible combinations of the form $\theta (DF)^2$ either vanish or
reduce to  $\theta F^3.$ Further, as  $\epsilon^{\m\n\r\s}$ does
not appear, only two invariants,
$\theta^{\m\n}F_{\m\n}F_{\r\s}F^{\r\s}$ and $ \theta^{\m\n}
F_{\m\r}F_{\n\s}F^{\r\s} $,  are left  at the end of the
calculation. Note that they are already present in the classical
action (\ref{actionexp}).

 The procedure to calculate the
divergencies in the traces is standard: first we write the traces
in the momentum representation,  then find their divergent parts
by dimensional regularization. An example of this calculation is
given in the Appendix. In the result we get:

{\setlength\arraycolsep{2pt}\bea
\Tr[(\Box^{-1}N_1)^2\Box^{-1}T_4]& = & \dfrac{i}{(4\pi
)^2\epsilon} d^{abc}\int
d^4x[2(\theta^{\r\a}F_{\m\a}^b+\theta_{\m\a}F^{\r\a b})(A^\m A_\n
A^\n A_\r)_{bc} \nonumber \\
&+& \frac12\theta^{\r\s}F_{\r\s}^b(A^\m A_\n A^\n
A_\m)_{bc}],\nonumber \eea} {\setlength\arraycolsep{2pt}\bea
\Tr[(\Box^{-1}N_2)^2\Box^{-1}T_2]&=&\dfrac{i}{(4\pi)^2
\epsilon}d^{abc}\int d^4
x\left[\frac{3}{2}\theta^{\r\s}F_{\r\s}^a(A_\m A^\m A_\n
A^\n)_{bc}\right.\nonumber \\
&-&\left.2({\theta^{\r}}_\a F_{\r\b}^a+ {\theta^{\r}}_\b
F_{\r\a}^a)(F^{\b\g}{F_\g} ^{\a})_{bc}+ \frac{1}{2} \theta^{\r\s}
F_{\r\s}^a (F_{\a\g}F^{\a\g})_{bc}\right],\nonumber \eea}
{\setlength\arraycolsep{2pt}\bea
\Tr[(\Box^{-1}N_1)^4\Box^{-1}T_2]&=& \dfrac{i}{(4\pi
)^2\epsilon}d^{abc}\int d^4 x\left[\frac{2}{3} \theta^{\r\s}
F_{\r\s}^a
(A_\a A^\a A_\b A^\b +A_\a A_\b A^\a A^\b\right. \nonumber \\
&+& \left. A_\a A_\b A^\b A^\a)_{bc} + (\theta_\r^{\ \a} F_{\s\a}^a
+ \theta_\s^{\ \a} F_{\r\a}^a)\left(\frac{2}{3} A_\m
A^\m A^\r A^\s\nonumber \right.\right.\\
&+&\left.\left. \frac{2}{3} A_\m A^\r A^\m A^\s+ \frac{1}{3} A_\m
A^\s A^\r A^\m + \frac{1}{3} A^\r A_\m A^\m A^\s\right)_{bc}
\right],\nonumber \eea}
{\setlength\arraycolsep{2pt}\bea\Tr[(\Box^{-1}N_1)^3\Box^{-1}T_3]&=&
\dfrac{i}{(4\pi)^2\epsilon} d^{abc}\int d^4 x \left[\frac{4}{3}
(\theta^{\r\a} F_{\s\a}^a +
\theta_{\s\a} F^{\r\a a}) (A_\r A^\s A^\m A_\m \right.\nonumber \\
&+& A_\r A_\m A^\s A^\m+A_\r A^\m A_\m A^\s)_{bc}\nonumber \\
&+&\left.\frac{1}{3}\theta^{\r\s}F_{\r\s}^a(A_\m A^\m A_\n A^\n
+A_\m A_\n A^\m A^\n+ A_\m A_\n A^\n A^\m)_{bc}\right],
\nonumber\eea}
{\setlength\arraycolsep{2pt}\bea\Tr[\Box^{-1}N_2(\Box^{-1}N_1)^2
\Box^{-1}T_2& +& (\Box^{-1}N_1)^2\Box^{-1}N_2\Box^{-1} T_2
+\Box^{-1} N_1 \Box^{-1} N_2 \Box^{-1} N_1 \Box^{-1} T_2]\nonumber
\\ & =& \dfrac{i}{(4\pi)^2\epsilon} d^{abc} \int d^4x
\left[\frac{7}{6}\theta^{\r\s}F_{\r\s}^a(2A_\m A^\m A_\n A^\n
\right.\nonumber\\ &+& A_\m A_\n
A^\n A^\m)_{bc}\nonumber \\
&+&
\left.\frac{2}{3}\left(\theta^{\r\a}F_{\s\a}^a+\theta_{\s\a}F^{\r\a
a})(A_\r A_\m A^\m A^\s +2 A_\r A^\s A_\m A^\m\right)_{bc}\right],
\nonumber\eea}
 {\setlength\arraycolsep{2pt}\bea
 \Tr[\Box^{-1}N_2\Box^{-1}T_4]&=&\dfrac{i}{(4\pi )^2\epsilon}
d^{abc}\int d^4x [4 \theta^{\r\a}F_{\m\a}^a (A^\m A^\b A_\b
A_\r)_{bc}\nonumber \\&+&\dfrac{1}{2} \theta^{\r\s}F_{\r\s}^a(A_\n
A_\m A^\m A^\n)_{bc} + 4i
\theta_{\r\a} F_{\m\b}^a (A^\m F^{\a\b} A^\r)_{bc} \nonumber \\
&-&{i} \theta^{\r\s}F_{\r\s}^a (A^\a F_{\a\b} A^\b)_{bc}
 -\dfrac{3N}{2}
\theta^{\r\s} F_{\r\s}^a F_{\a\b}^b F^{\a\b c}\nonumber \\
&-&  4i\theta^{\r\a} F_{\m\r}^a (A^\m F_{\a\b} A^\b)_{bc}
 +4 i\theta^{\r\s} F_{\b\r}^a (A_\a F^{\b\a}
A_\s)_{bc}\nonumber \\
&-& 2i \theta^{\a\b} F_{\m\n}^a (A^\n F_{\a\b} A^{\m})_{ab}-2i
\theta_{\a\r} F_{\m\b}^a (A^\r F^{\b\a} A^\m)_{bc} \nonumber \\
&-& 2 i\theta_{\b\s} F_{\m\a}^a(A^\m F^{\b\a} A^\s)_{bc} +2 i
\theta^{\r\s} F_{\a\b}^a (A_\s F^{\b\a} A_\r)_{bc}\nonumber \\
&+&2 N \theta^{\r\s}F_{\a\r}^a F_{\b\s}^b  F^{\a\b c} -4N
\theta^{\r\b} F_{\m\r}^a F^{\m\a b}F_{\b\a}^c ],\nonumber \eea} and
{\setlength\arraycolsep{2pt}\bea
\Tr(\Box^{-1}N_1\Box^{-1}N_2\Box^{-1}T_3&+&\Box^{-1}N_2\Box^{-1}N_1\Box^{-1}T_3)
\nonumber \\
& =&\dfrac{i}{(4\pi )^2\epsilon}d^{abc}\int d^4
x[\dfrac12\theta^{\r\s}F_{\r\s}^a(A_\a A^\a A_\b A^\b \nonumber
\\ &+& A_\m A_\a A^\a A^\m)_{bc} + 2\theta^{\r\s}F_{\m\s}^a(A^\m
A_\r A_\a A^\a \nonumber \\ &+& A_\r A^\m A_\a
A^\a +2A^\m A_\a A^\a A_\r)_{bc}\nonumber \\
&+&\dfrac{1}{2}\theta^{\r\s}F_{\r\s}^a(F_{\a\b}F^{\a\b})_{bc}
+4\theta_{\r\a}F_{\m\b}^a(F^{\r\m}F^{\a\b})_{bc}\nonumber\\
&-&i\theta^{\r\s}F_{\r\s}^a(A^\a F_{\a\b}A^\b)_{bc}
+8i\theta_{\r\a}F_{\m\b}^a(A^\m F^{\a\b}A^\r)_{bc} \nonumber\\
&-&2\theta^{\r\a}F_{\m\r}^a(F^{\b\m}F_{\a\b})_{bc}- 2
\theta^{\r\s} F_{\b\r}^a (F_{\s\a}F^{\a\b})_{bc}\nonumber\\
&+&\theta^{\a\b}F_{\r\n}^a(F^{\n\r}F_{\a\b})_{bc}
+\theta^{\r\s}F_{\a\b}^a(F_{\s\r}F^{\a\b})_{bc}\nonumber\\
&-&4i\theta^{\r\a}F_{\m\r}^a(A^\m F_{\a\b}A^\b)_{bc}
-4i\theta^{\r\s}F_{\b\r}^a(A_\a F^{\a\b}A^\s)_{bc}\nonumber\\
&-&2i\theta^{\a\b}F_{\m\n}^a(A^\n F_{\a\b}A^\m)_{bc}
-2i\theta^{\r\s}F_{\a\b}^a(A_\s F^{\a\b}A_\r)_{bc}] . \nonumber
\eea}

The difficult part, after summation of particular diagrams, is to
transform the sum to the covariant form. We obtain
 \beq  \Gamma^{1}|_{\rm
div}=-\frac{11N}{6(4\pi )^2\epsilon}d^{abc}\int
d^4x[\frac14\theta^{\r\s}F_{\r\s}^aF_{\m\n}^bF^{\m\n
c}-\theta^{\r\s}F_{\r\m}^aF_{\s\n}^bF^{\m\n c}]  , \eeq where at
the end of the calculation we apply the formula from \cite{mcF}, namely
 {\setlength\arraycolsep{2pt}\bea
d^{abc}F_{\a\b}^a(F_{\m\n}F_{\r\s})_{bc}&=&
d^{abc}F_{\a\b}^aF_{\m\n}^d
F_{\r\s}^e(T^dT^e)_{bc}\nonumber \\
&=& F_{\a\b}^aF_{\m\n}^dF_{\r\s}^e\Tr(D^aT^dT^e) \nonumber\\ &=&
\dfrac{N}{2}d^{ade}F_{\a\b}^aF_{\m\n}^dF_{\r\s}^e , \nonumber
\eea} \noindent
with $(D^a)_{bc}=d^{abc}$. This is the only place
where the assumption that  fields are in the adjoint
representation is used. The full result for the one-loop divergent
part of the effective action to first order  is
{\setlength\arraycolsep{2pt}\bea \G_{\rm
div}&=&-\frac14\left(1-\frac{22N}{3(4\pi)^2\epsilon}\right)\int
d^4xF_{\m\n}^aF^{\m\n a}\nonumber\\&+&\frac14\left(1-
\frac{22N}{3(4\pi)^2\epsilon}\right)\theta^{\r\s}d^{abc}\int
d^4x\left(\frac14F_{\r\s}^aF_{\m\n}^bF^{\m\n
c}-F_{\m\r}^aF_{\n\s}^bF^{\m\n c}\right)                 .
\label{G} \eea} From the last equation it is obvious that  the
one-loop correction to the effective action is proportional to the
classical action. The (\ref{G}) can be rewritten in the form \beq
\G_{\rm
div}=-\frac14\left(1-\frac{22N}{3(4\pi)^2\epsilon}\right)\int d^4x
F_{\m\n}^a\star F^{\m\n a}. \eeq

\initiate
\section{Outlook and conclusions}

In the presented calculation  the coupling constant was fixed to be 1.
However, in order to renormalize the theory, we have to recover it.
 We take that, initially, the classical
lagrangian in $4-\epsilon$ dimensions reads
\begin{equation}
\cl = -\frac{1}{4}F_{\m\n }^aF^{\m\n a}+
\frac{1}{16}g\mu^{\epsilon
/2}\theta^{\m\n}d^{abc}(F_{\m\n}^aF_{\r\s}^bF^{\r\s c}
-4F_{\m\r}^aF_{\n\s}^bF^{\r\s c}),
\end{equation}
while the field strength is defined by \beq F_{\m\n}^a = \pa
_\m A_\n^a -\pa_\n A_\m^a +g\m^{\e/2} f^{abc}A_\m^b A_\n^c . \eeq
The $\m$ is  a parameter with the dimension of mass. To cancel
divergencies we add counterterms to the initial action. The bare
Lagrangian is the sum of the classical Lagrangian and the
counterterms; it reads {\setlength\arraycolsep{2pt}\bea \cl _0&=&
\cl+\cl_{CT} \label{L} \\
&=&-\frac14\left(1+\frac{22Ng^2}{3(4\pi)^2\epsilon}\right)
F_{\m\n}^aF^{\m\n a}\nonumber\\&+&\frac14\left(1+
\frac{22Ng^2}{3(4\pi)^2\epsilon}\right)\theta^{\r\s}d^{abc}g\mu
^{\epsilon /2}\left(\frac14F_{\r\s}^aF_{\m\n}^bF^{\m\n
c}-F_{\m\r}^aF_{\n\s}^bF^{\m\n c}\right)                 . \nonumber
\eea}
Introducing the bare quantities
 \bea
{A_0}^{\m a}&=&A^{\m a}\sqrt{1+\frac{22Ng^2}{3(4\pi)^2\epsilon}},\nonumber\\
g_0&=&\frac{g\m^{\e/2}}{\sqrt{1+\frac{22Ng^2}{3(4\pi)^2\epsilon}}},\label{g}
 \eea
we can rewrite the bare Lagrangian as
\begin{equation}
\cl_0 = -\frac{1}{4}{F_0}_{ \m\n }^a{F_0}^{\m\n a}+
\frac{1}{16}g_0\theta^{\m\n}d^{abc}({F_0}_{\m\n}^a{F_0}_{\r\s}^b{F_0}^{\r\s
c} -4{F_0}_{\m\r}^a{F_0}_{\n\s}^b{F_0}^{\r\s c}),                             \label{Lbare}
\end{equation}
or, \beq \cl=-\frac14 {F_0}_{\m\n}^a\star {F_0}^{\m\n a}. \eeq
Note that, due to the fact that divergencies of kinetic  and
$\theta$-linear interaction terms have the same factor,
noncommutativity parameter $\theta$ need not be renormalized. This
rises hope that the theory might be renormalizable to all orders in
$\theta$. One can easily find the beta function from (\ref{g}): it
is same as in the commutative case \beq \beta =\m\frac{\pa g}{\pa
\m}= -\frac{11Ng^3}{3(4\pi)^2}.\eeq The theory is asymptotically
free.

Formulae (\ref{g}-\ref{Lbare}) mean that the noncommutative pure
gauge $SU(N)$ theory is one-loop renormalizable to  first order in
$\theta$. Divergences can be absorbed in the redefinition of the
gauge potential and the gauge coupling constant. The
renormalization is standard, multiplicative: no Seiberg-Witten
field redefinition is needed, as it was the case in  similar
calculations \cite{mv,u1}.  A different result  would further
strengthen the belief that field theories on noncommutative
Minkowski space are  not renormalizable. As it is, it opens
several possibilities. The first possibility is that the gauge
theories are renormalizable in the $\theta$-expanded approach
because in this representation the gauge symmetry is introduced in
a natural way, via the covariant coordinates. Then fermions are
probably inadequately represented, as we know that their presence
breaks renormalizability. An  obvious further step to check this
claim is, for example, to find the second-order divergencies;
also, one could consider renormalizability at two loops. The other
possible interpretation is that renormalizability is obstructed by
the use of
 Moyal-Weyl $\star$-product
\cite{madore}, and that the more appropriate representation has to
be found.  In any case, the question deserves further
consideration.

\vskip0.5cm {\bf Acknowledgment} \ \ This work is supported by the
Serbian Ministry of Science Grant No. 1486.

\vskip1cm

\initiate \begin{Large}{\bf Appendix: an example of trace
calculation}\end{Large} \vskip0.5cm

In this appendix we present the calculation of the diagram
$\Tr((\Box^{-1}N_1)^4\Box^{-1}T_2)$; other terms in the one-loop
effective action are computed in a similar manner.
{\setlength\arraycolsep{2pt}\bea
\Tr((\Box^{-1}N_1)^4\Box^{-1}T_2)&=&\int d^4x d^4y d^4z d^4u d^4v
\ g^{\a\kappa}G(x-y){N_1}_{\a\b}^{ a b}(y)
\nonumber\\&\times&G(y-z){N_1}^{\b\g bc}(z)G(z-u){N_1}_{\g \d
}^{cd}(u)\nonumber \\ &\times& G(u-v){N_1}^{\d\lambda d
e}(v)G(v-x){T_2}_{\lambda \kappa }^{ea}(x)\nonumber\\&=&
16d^{abc}\int \frac{d^4x}{(2\pi)^4}\int d^D p\frac{p^\a p^\b p^\m
p^\n p^\r p^\s }{(p^2)^5}\nonumber\\&\times&(A_\a A_\b A_\m A_\n
)_{bc}(2\theta^{\s}_{\ \e}F^{\r\e a}+\frac14
g^{\r\s}\theta^{\e\l}F_{\e\l}^a).\eea}
 All external momenta in
vertices vanish, as a consequence of (\ref{Aconst}). The divergent
part of the previous integral is found by dimensional
regularization. The result is {\setlength\arraycolsep{2pt}\bea
\Tr((\Box^{-1}N_1)^4\Box^{-1}T_2)&=&d^{abc}\frac{i}{6(4\pi)^2\e}\int
d^4x(A_\a A_\b A_\m A_\n )_{bc}\nonumber\\&\times&(2\theta^{\s}_{\
\e}F^{\r\e a}+\frac14
g^{\r\s}\theta^{\e\l}F_{\e\l}^a)\nonumber\\
&\times&[g_{\a\b}(g_{\m\n}g_{\r\s}+g_{\m\r}g_{\n\s}+g_{\m\s}g_{\n\r})\nonumber\\
&+&g_{\a\m}(g_{\b\n}g_{\r\s}+g_{\b\r}g_{\n\s}+g_{\b\s}g_{\n\r})\nonumber\\
&+&g_{\a\n}(g_{\b\m}g_{\r\s}+g_{\b\r}g_{\m\s}+g_{\b\s}g_{\m\r})\nonumber\\
&+&g_{\a\r}(g_{\b\m}g_{\n\s}+g_{\b\n}g_{\m\s}+g_{\b\s}g_{\n\m})\nonumber\\
&+&g_{\a\s}(g_{\b\m}g_{\n\r}+g_{\b\n}g_{\m\r}+g_{\b\r}g_{\m\n})]\
.\nonumber\eea} From the previous expression we get the final
result \bea \Tr[(\Box^{-1}N_1)^4\Box^{-1}T_2]&=& \dfrac{i}{(4\pi
)^2\epsilon}d^{abc}\int d^4 x\left[\frac{2}{3} \theta^{\r\s}
F_{\r\s}^a
(A_\a A^\a A_\b A^\b +A_\a A_\b A^\a A^\b\right. \nonumber \\
&+& \left. A_\a A_\b A^\b A^\a)_{bc} + (\theta_\r^{\ \a}
F_{\s\a}^a + \theta_\s^{\ \a} F_{\r\a}^a)\left(\frac{2}{3} A_\m
A^\m A^\r A^\s\nonumber \right.\right.\\
&+&\left.\left. \frac{2}{3} A_\m A^\r A^\m A^\s+ \frac{1}{3} A_\m
A^\s A^\r A^\m + \frac{1}{3} A^\r A_\m A^\m A^\s\right)_{bc}
\right]\ . \nonumber \eea


\begin{thebibliography}{99}

\bibitem{phi4} S.~Minwalla, M.~Van Raamsdonk and N.~Seiberg,
JHEP {\bf 0002} (2000) 020; I.~Chepelev and R.~Roiban, JHEP {\bf
0005} (2000) 037; JHEP {\bf 0103} (2001) 001;
 C.~Becchi, S.~Giusto and C.~Imbimbo,
 Nucl.Phys. B {\bf 664} (2003) 371;
 H.~Grosse and R.~Wulkenhaar, Commun. Math. Phys. {\bf 256} (2005) 305

\bibitem{Gay} V.~Gayral, J.~M.~Gracia-Bondia and F.~Ruiz~Ruiz,
Phys. Lett. B {\bf 610} (2005) 141

\bibitem{U1}
M.~Hayakawa,
Phys.Lett. B {\bf 478} (2000) 394;
 C.~P.~Martin and D.~Sanchez-Ruiz, Phys.Rev. Lett. {\bf 83} (1999) 476;
I.~F.~Riad and M.~M.~Sheikh-Jabbari,
JHEP {\bf 0008} (2000) 045

\bibitem{nonab} A.~Matusis, L.~Susskind and N.~Toumbas,
 JHEP {\bf 0012} (2000) 002;
L.~Bonora and M.~Salizzoni, Phys. Lett. B {\bf 504} (2001) 80; A.~Armoni,
 Nucl. Phys. B {\bf 593} (2001) 229;
M.~Van Raamsdonk, JHEP {\bf 0111} (2001) 006;
A.~Armoni and E.~Lopez,
  Nucl. Phys. B {\bf 632} (2002) 240; E.~Nicholson,
Phys.Rev. D {\bf 66} (2002) 105018

\bibitem{thetaexp} J.~Madore, S.~Schraml, P.~Schupp and J.~Wess, Eur.
Phys. J. C {\bf 16} (2000) 161; B.~Jur\v co, S.~Schraml, P.~Schupp
and J.~Wess,  Eur. Phys. J. C {\bf 17} (2000) 521; B.~Jur\v co, L.
M\"oller, S.~Schraml, P.~Schupp and J.~Wess,  Eur. Phys. J. C {\bf
21} (2001) 383

\bibitem{SW} N.~Seiberg and E.~Witten, JHEP {\bf 9909} (1999) 032

\bibitem{a} T.~Asakawa and I.~Kishimoto,  JHEP {\bf 9911} (1999) 024

\bibitem{w} R.~Wulkenhaar,  JHEP {\bf 0203} (2002) 024

\bibitem{mv} M.~Buri\' c and V.~Radovanovi\' c, JHEP {\bf 0210} (2002) 074

\bibitem{mv1} M.~Buri\' c and V.~Radovanovi\' c, JHEP {\bf 0402} (2004) 040;
M.~Buri\' c and V.~Radovanovi\' c, Class. Quant. Grav. {\bf 22} (2005) 525

\bibitem{u1} A.~A.~Bichl, J.~M. Grimstrup, H.~Grosse, L.~Popp, M.~Schweeda
and R.~Wulkenhar, JHEP {\bf 0106} (2001) 013

\bibitem{thoft} G.~'t~Hooft, Nucl. Phys. B {\bf 62} (1973) 444

\bibitem{weinberg} S. Weinberg,
 {\it The Quantum Theory of Fields}, vol. II, Cambridge University Press, New York, 1996

\bibitem{ps} M.~E.~Peskin and D.~V.~Schroeder, {\it An introduction to
Quantum Field Theory} Addison Wesley, Reading, 1995

\bibitem{mcF} J.~A.~de Azcarraga, A.~J.~Macfarlane, A.~J.~Mountain and J.~C.~Perez
Bueno, Nucl. Phys. B {\bf 510} (1998) 657

\bibitem{madore} J.~Madore, private communication

\end{thebibliography}
\end{document}